\documentclass[aps,prl,reprint,superscriptaddress,showpacs]{revtex4-1}

\usepackage{graphicx}			
\usepackage{bm}				
\usepackage{amsmath}			
\usepackage{amsthm}			
\usepackage{amssymb}			
\usepackage{multirow}

\begin{document}

\title{Topological Classification of Crystalline Insulators with Point Group Symmetry}

\author{Priyamvada Jadaun}
\affiliation{Microelectronics Research Center, The University of Texas at Austin, Austin, TX 78758}

\author{Di Xiao}
\email{dixiao@cmu.edu}
\affiliation{Department of Physics, Carnegie Mellon University, Pittsburgh, PA 15213}

\author{Qian Niu}
\affiliation{Department of Physics, The University of Texas at Austin, Austin, TX 78712}

\author{Sanjay K. Banerjee}
\affiliation{Microelectronics Research Center, The University of Texas at Austin, Austin, TX 78758}

\date{\today}

\begin{abstract}
We show that in crystalline insulators point group symmetry alone gives rise to a topological classification based on the quantization of electric polarization.  Using $C_3$ rotational symmetry as an example, we first prove that the polarization is quantized and can only take three inequivalent values.  Therefore, a $Z_3$ topological classification exists.  A concrete tight-binding model is derived to demonstrate the $Z_3$ topological phase transition.  Using first-principles calculations, we identify graphene on BN substrate as a possible candidate to realize the Z3 topological states.  To complete our analysis we extend the classification of band structures to all 17 two-dimensional space groups.  This work will contribute to a complete theory of symmetry conserved topological phases and also elucidate topological properties of graphene like systems. 
\end{abstract}

\maketitle

Since the celebrated discovery of the quantum Hall effect~\cite{qhe}, topological classification of electronic states has emerged as a powerful concept in condensed matter physics.  The quantum Hall insulators are distinguished from ordinary insulators by a topological index, the TKNN number, which gives the quantized Hall conductance~\cite{thouless,niu}. For a long time, the TKNN number was thought to be the only topological index describing non-degenerate electronic ground states.  Recently, it was realized that in crystalline insulators new topological indices can be defined in the presence of discrete symmetries, which has led to the identification of a slew of new topological states.  For example, the quantum spin Hall insulators are characterized by a nontrivial $Z_2$ index~\cite{kane}, which is protected by time-reversal symmetry~\cite{hasan1,xlqi1}.  Similarly, magnetic translation symmetry can also give rise to a $Z_2$ classification in antiferromagnetic insulators~\cite{moore2}.  Another interesting proposal is the so-called topological crystalline insulators, in which a $Z_2$ index can be defined and is protected by both time-reversal and certain point group symmetries~\cite{liangfu,bansil}.  

In this Letter we show that in crystalline insulators point group symmetry \emph{alone} can give rise to a new topological classification based on the quantization of electric polarization.  Our idea is inspired by a beautiful result due to Zak~\cite{zak}, i.e., in one-dimensional (1D) systems with inversion symmetry, the Berry phase of the Bloch bands can be either 0 or $\pi$~\cite{xiao}.  This quantization of the Berry phase naturally leads to a $Z_2$ classification in 1D.  In higher dimensions, we find that the role of Zak's phase is replaced by a closely related quantity, the electric polarization~\cite{vanderbilt,resta}, which is quantized in the presence of point group symmetry.  The generalization to higher dimensions is expected to display a richer spectrum of possibilities because of the enlarged symmetry class compared to 1D.

For the sake of definiteness, 2D crystals with $C_3$ rotational symmetry are used as an example in the following discussion.  We first prove, from general symmetry argument, that the polarization is quantized and can only take three inequivalent values.  Therefore, a $Z_3$ topological classification exists.  We then proceed by constructing a concrete tight-binding model to demonstrate the existence of various $Z_3$ topological states and the topological phase transition in the presence of $C_3$ symmetry.  Using first-principles calculations, we identify graphene on BN substrate~\cite{giovannetti,bn,dean,ci} as a possible candidate to realize the $Z_3$ topological states.  The misconception that $C_3$ symmetry enforces zero polarization is also clarified.  Finally, we extend our analysis to all space groups in 2D.  Since a many-body formulation of the electric polarization already exists~\cite{martin,resta2}, the topological classification scheme discussed here can be readily applied to interacting systems.

\textit{General symmetry analysis.}---Theoretically, it has been well established that the polarization in a crystal is formally defined only modulo a quantum uncertainty~\cite{vanderbilt,resta}, i.e., the polarization vectors $\vec P$ given in the following expression are all equivalent to each other:
\begin{equation}
\label{eq1}
\vec P \simeq \vec P + e\sum_i n_i \vec a_i \;,
\end{equation}
where $e$ is the electron charge, $n_i$ are integers, and  $\vec a_i$ are the primitive lattice vectors.  This is a direct consequence of the translation symmetry of the lattice.  At the same time, $\vec P$ has to also satisfy the point group symmetry of the crystal, which places great restriction on the allowed values of $\vec P$, thereby quantizing it.  For example, in 1D the only point group symmetry is inversion, which requires that $\vec P$ and $-\vec P$ must be equivalent.  This leads to two possibilities of $\vec P$ (0 and $ea/2$) and Zak's phase (0 and $\pi$)~\cite{zak}, and a $Z_2$ classification.

Let us now consider a 2D lattice with $C_3$ rotational symmetry.  We denote the two primitive lattice vectors by $\vec a$ and $\vec b$.  The polarization vector can be expressed as $\vec P = e(\alpha \vec a + \beta \vec b)$.  Consider the in-plane rotation $\hat R$ by an angle of $2\pi/3$.  $\hat R$ operates on the lattice vectors such that
\begin{equation}
\hat R(\vec a) = \vec b - \vec a \;, \quad
\hat R(\vec b) = -\vec a \;,
\end{equation}
and $\vec P$ transforms according to
\begin{equation} \label{pp}
\hat R(\vec P) = \vec P' = e(\alpha \vec b - (\alpha + \beta) \vec a) \;.
\end{equation}
However, the transformed polarization vector $\vec{P'}$ must be equivalent to the starting value $\vec P$ according to Eq.~\eqref{eq1}, i.e.,
\begin{equation}
\vec P' = \vec P  + e (m\vec a + n\vec b ) \;.
\end{equation}
Combining the above equations of $\vec P'$ yields possible values for $\alpha$ and $\beta$:
\begin{equation}\label{eq2}
\alpha = \frac{n-m}{3} = \frac{j}{3}\;, \quad
\beta = -\frac{m+2n}{3} = \frac{j}{3} - n \;.
\end{equation}
where $n$, $m$ and $j$ are integers.  The polarization values allowed by $C_3$ symmetry thus fall into 3 inequivalent sets described by 0, $(\vec a + \vec b)/3$, $2(\vec a + \vec b)/3$, and corresponding series of equivalent values given by Eq.~\eqref{eq1}.

This result has two important consequences.  First, it shows that while zero is an allowed value, $C_3$ symmetry does not preclude nonzero polarization values as is sometimes assumed in the literature.  This misconception is a result of the classical intuition that like forces, one can cancel polarization vectors in opposing directions.  While this may apply for dipole vectors in finite, discrete charge distributions, it is not true for bulk crystalline insulators. Secondly, and more importantly, the above result suggests that 2D crystalline insulators with $C_3$ symmetry can be distinguished by the three possible values of $\vec P$, giving rise to a $Z_3$ classification.  Following the same procedure, one can verify that $C_6$ rotational symmetry, e.g., as seen in graphene, allows only zero polarization in insulators.

\textit{$Z_3$ Topological phase transition.}---We now provide a concrete example of $Z_3$ topological states by constructing a tight-binding model.  Let us consider a honeycomb lattice with two orbitals, denoted by $\phi_\alpha$ and $\phi_\beta$, on each site.  We assume that there is a nearest-neighbor hopping $t$ between orbitals from the same type, and an on-site inter-orbital coupling $t'$ between the $\alpha$ and $\beta$ orbitals.  To lower the symmetry from $C_6$ to $C_3$, the $A$ and $B$ sublattices are assumed to have opposite site energies, $\pm\Delta_\alpha$ and $\pm\Delta_\beta$, for each type of orbitals.  The tight-binding Hamiltonian is given by
\begin{equation}
\begin{split}
H &= t\sum_{\langle ij\rangle} (c^\dag_{\alpha i} c_{\alpha j} + c^\dag_{\beta i} c_{\beta j}) + t' \sum_i (c^\dag_{\alpha i} c_{\beta i} + c^\dag_{\beta i} c_{\alpha i}) \\
& \quad + \sum_i \xi_i (\Delta_\alpha c^\dag_{\alpha i} c_{\alpha i} +\Delta_\beta c^\dag_{\beta i} c_{\beta i}) \;,
\end{split}
\end{equation}
where $\xi_i = \pm 1$ for A and B sublattices.  The corresponding Bloch Hamiltonian is obtained by Fourier transform:
\begin{equation} \label{tb}
H(\vec k) = \begin{pmatrix} 
\Delta_\alpha & t V_k & t' & 0 \\
t V^*_k & -\Delta_\alpha & 0 & t' \\
t' & 0 & \Delta_\beta & t V_k \\
0 & t' & t V^*_k & -\Delta_\beta 
\end{pmatrix} \;,
\end{equation} 
where the structural factor $V_k$ is given by $V_k = 1 + 2\cos(\sqrt{2}/2 k_x)\cos(k_y/2)-2i\sin(\sqrt{3}k_x/2)\cos(k_y/2)$.  

\begin{figure}
\includegraphics[width=6cm]{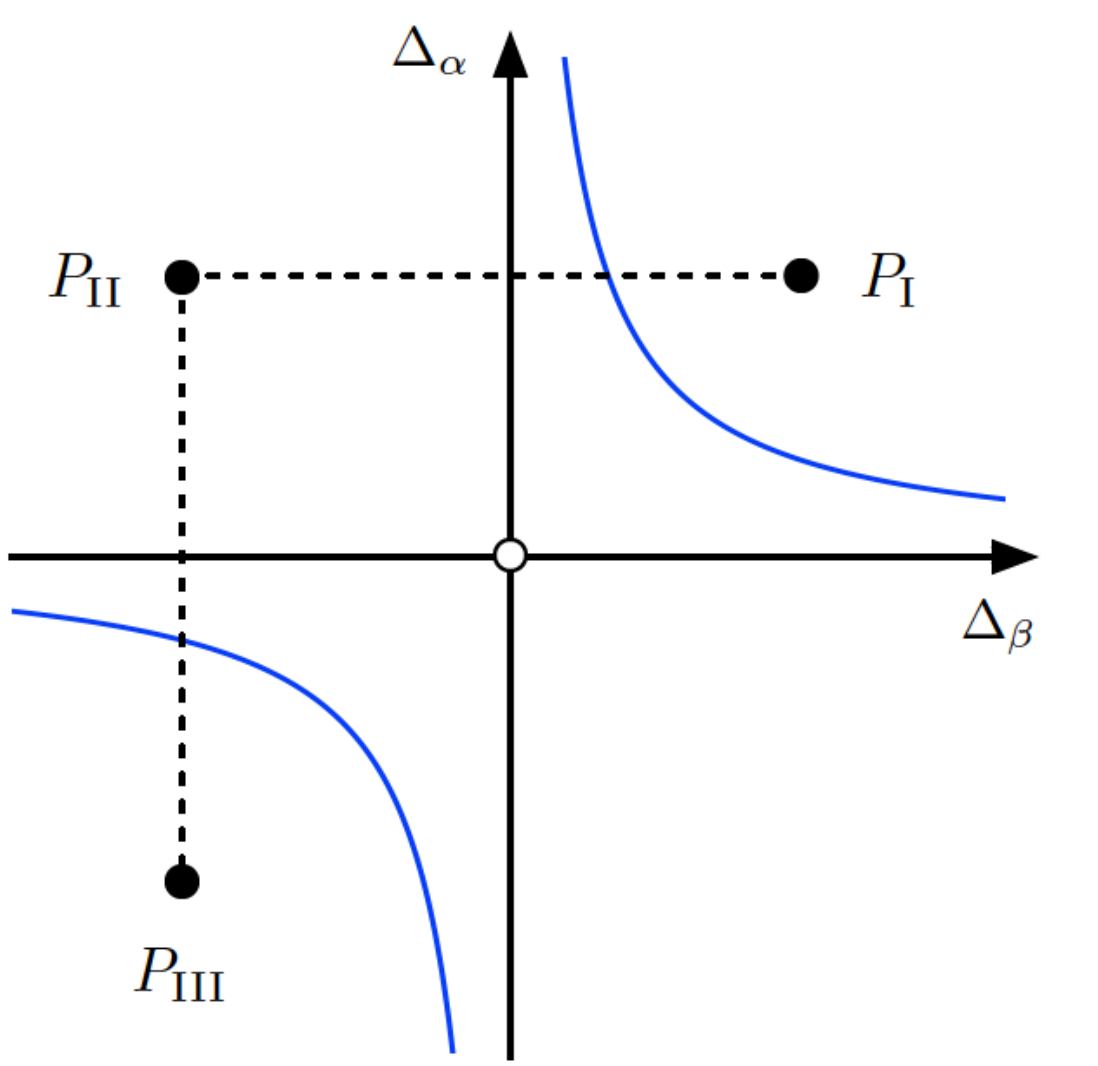}
\caption{\label{diagram} (color online) Topological phase diagram of the tight-binding model described by the Hamiltonian~\eqref{tb}.  The solid (blue) curve is the phase boundary given by $\Delta_\alpha\Delta_\beta = t'^2$.  In each region, the polarization is given by $P_\text{I} = -(\vec a + \vec b)/3$, $P_\text{II} = 0$, and $P_\text{III} = (\vec a + \vec b)/3$.  The dashed line indicates a representative path in the parameter space that undergoes the topological phase transition twice.}
\end{figure}

Next we show that at half-filling the system can be tuned to three topologically distinctive states by changing the parameters without breaking the $C_3$ symmetry.  We first identify the gap closing point in the parameter space because $\vec P$ cannot change unless the band gap closes.  The band energies at the $K$ point is given by
\begin{equation}
E = \pm(\Delta_\alpha + \Delta_\beta) \pm \sqrt{(\Delta_\alpha - \Delta_\beta)^2 + 4t'^2} \;.
\end{equation}
Obviously, when $t' = \sqrt{\Delta_\alpha\Delta_\beta}$ the band gap will close, which signals a topological phase transition characterized by a sudden change of $\vec P$.  To confirm this statement, we calculate the polarization according to the Berry phase formula~\cite{vanderbilt,resta} at three representative points in the $\Delta_\alpha-\Delta_\beta$ plane, and indeed find three inequivalent values of $\vec P$ predicted by Eq.~\eqref{eq2}.  The resulting topological phase diagram is shown in Fig.~\ref{diagram}.

\textit{Material realization.}---We now present a realistic material system, graphene on $h$-BN substrate~\cite{giovannetti,bn,dean,ci}, in which various $Z_3$ topological states can be realized.  Specifically, we consider four cases: $h$-BN, AA stacking of graphene on $h$-BN, and AB stacking of graphene on $h$-BN with either boron or nitrogen atoms sitting directly underneath the carbon atom.  Although graphene has $C_6$ rotational symmetry, placing graphene on $h$-BN reduces the symmetry from $C_6$ to $C_3$, allowing non-zero polarization values to emerge. 

\begin{figure}[t]
\includegraphics[width=\columnwidth]{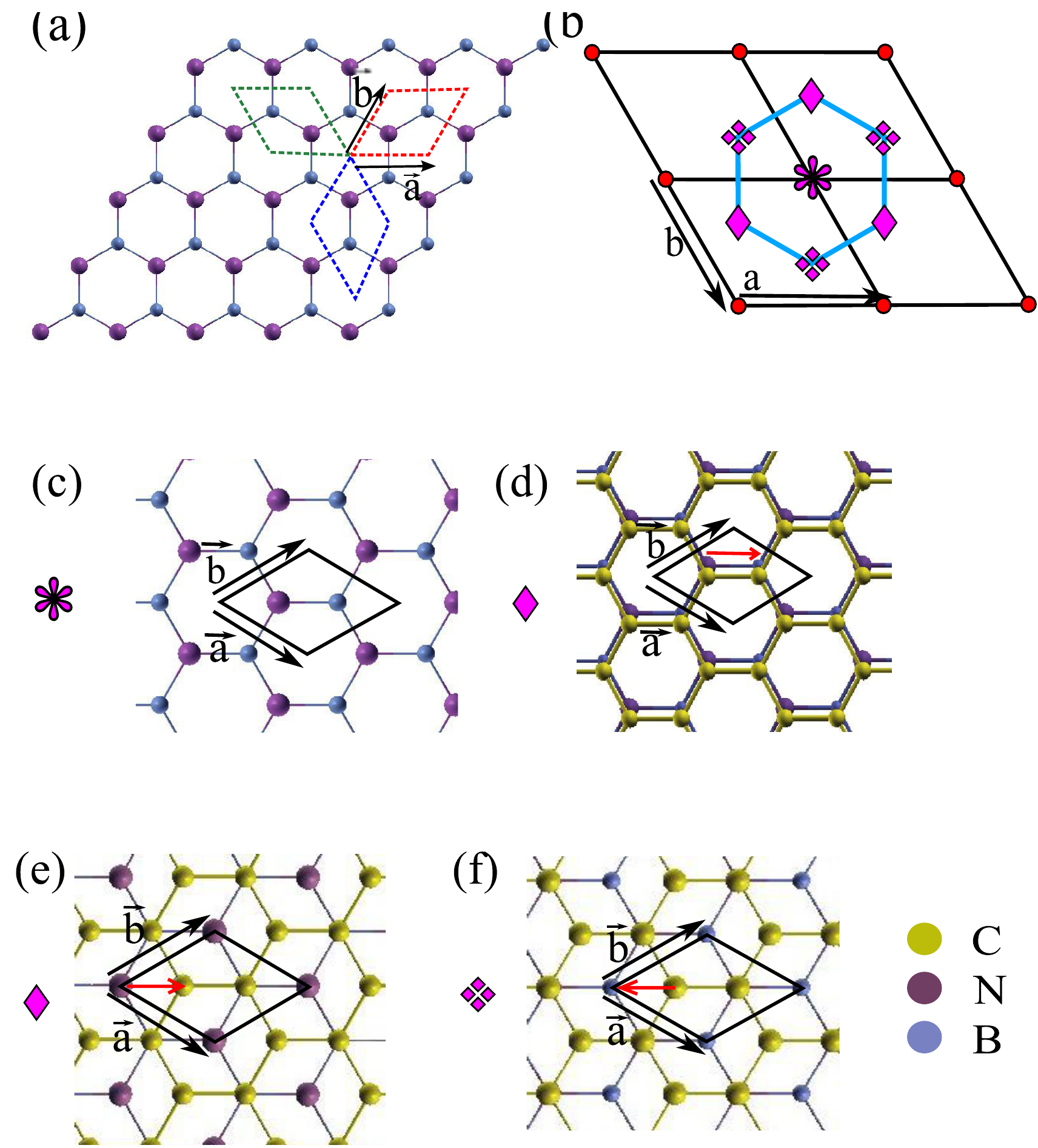}
\caption{(Color online) (a),(b) Schematic illustration of polarization in systems with $C_3$ symmetry.  (a) \textit{h}-BN lattice with 3 equivalent unit cells. (b) Possible values of polarization. Each set of topologically-equivalent values is marked by a specific symbol. (c)-(f) displays calculated values of polarization for different structures and the corresponding topological classes. (c) Structure 1: \textit{h}-BN monolayer; (d) Structure 2: AA stacking of graphene monolayer on \textit{h}-BN; (e) AB stacking of graphene monolayer on text{h}-BN with a carbon sublattice lying above the boron sublattice and (f) Structure 4: AB stacking with a carbon sublattice lying above the nitrogen sublattice. The unit cells and lattice vectors $\protect\vec{a}$ and $\protect\vec{b}$ are marked in black, whereas the polarization vector is marked in red. Colors used to represent various atoms are listed.\label{fig1}}
\end{figure}

We have performed first-principles calculations to obtain the polarization in these structures and compare with our symmetry analysis.  While the carbon-carbon (C-C) bond length in graphene is known to be 1.42{\AA}, \textit{h}-BN has a B-N bond length of 1.46{\AA}~\cite{cabrera}; hence the mismatch is very small $\sim2\%$.  Our initial structures comprised, as shown in Fig.\ref{fig1}, of bulk monolayer \textit{h}-BN and various stacking of graphene monolayer on it. We lattice matched \textit{h}-BN to graphene by taking an initial lattice constant of 1.417{\AA} and constructed super cells representing these materials in bulk (see Fig.\ref{fig1}). The out-of-plane distance between graphene and \textit{h}-BN was taken to be 3.2{\AA}~\cite{giovannetti}. The vacuum along the $z$-direction was taken to be 30{\AA} approximately. We used the Vienna Ab-initio Simulation Package (VASP)~\cite{kresse} which uses a plane-wave basis set and employed PAW pseudopotentials~\cite{paw}, along with a local approximation to the exchange-correlation potential (LDA)~\cite{perdewnzunger}. We set an energy cut off of 400 eV and a $k$-mesh of 4x4x1 for the self-consistent run and 36x36x1 for the subsequent polarization calculation.

The total polarization can be divided into the electronic and ionic parts.  While the electronic part of the polarization was calculated using the Berry phase formula~\cite{vanderbilt,resta}, the ionic part was obtained from the dipole density distribution. Polarization values thus calculated for bulk structures shown in Fig.\ref{fig1}, are listed in Table \ref{table1}, in corresponding order. We list $\alpha$ and $\beta$ for every $\vec{P}$, such that $\vec P = e(\alpha\vec a + \beta\vec b)$, where $\vec a$ and $\vec b$ are the corresponding lattice vectors of the system. In complete agreement with our symmetry analysis, all obtained values fell into and exhausted the 3 topological categories predicted by Eq.~\eqref{eq2}.  

\begin{table}[t] 
\caption{In-plane polarization values for the structures shown in Fig.\ref{fig1}, specified in terms of lattice vectors $\vec a$ and $\vec b$. We tabulate the values of the ionic part $\vec P_{ion}$ and the electronic part $\vec P_{ele}$, followed by the total value of polarization $\vec P$. All values are in terms of electronic charge$\times$lattice vector such that $\vec P = e(\alpha \vec a + \beta \vec b)$. \label{table1}}
\begin{ruledtabular}
\begin{tabular}{|c|c|c|c|c|c|c|}
System 	& \multicolumn{2}{|c|}{$\vec P_{ele}$}  &  \multicolumn{2}{|c|}{$\vec P_{ion} $} & \multicolumn{2}{|c|}{$\vec P $}\\
\hline
	& $\alpha$  &   $\beta$	   &      $\alpha$  &  $\beta$     &      $\alpha$   &  $\beta$\\
	&($e*\vec{a}$)&($e*\vec{b}$)&($e*\vec{a}$)&($e*\vec{b}$)&($e*\vec{a}$)&($e*\vec{b}$)\\
	\hline
1 	& $-\frac{2}{3}$ & $-\frac{2}{3}$ & $\frac{2}{3}$ & $\frac{2}{3}$ & 0 & 0 \\
2 	& 0 & 0 	& $\frac{1}{3}$ & $\frac{1}{3}$ & $\frac{1}{3}$ & $\frac{1}{3}$ \\
3 	& $\frac{2}{3}$ & $\frac{2}{3}$ & 0  & 0  & 		$\frac{2}{3}$ & $\frac{2}{3}$ \\
4 	& $\frac{2}{3}$ & $\frac{2}{3}$ & $\frac{2}{3}$ & $\frac{2}{3}$ & $\frac{1}{3}$  & $\frac{1}{3}$  \\
\end{tabular}
\end{ruledtabular}
\end{table}

Here we comment on the zero polarization found in $h$-BN~\cite{mele,mele2}.  The electronic structure of $h$-BN can be well described by taking the ionic limit, i.e., the boron atom loses all its valence electrons to nitrogen.  The nominal charge at the boron and nitrogen sites are therefore $+3$ and $-3$, respectively. (One can verify that in the ionic limit, the ionic and electronic contributions agree with the DFT calculation.)  This guarantees that the total polarization of $h$-BN will always be an integer multiple of $\vec a + \vec b$, which is equivalent to zero.  If we make an artificial crystal by replacing boron and nitrogen with beryllium and oxygen, then the total polarization will not vanish.  So the zero polarization of $h$-BN is not because $C_3$ rotational symmetry forbids other values, instead it is due to simple electron counting.

\begin{figure}[t]
\includegraphics[width=\columnwidth]{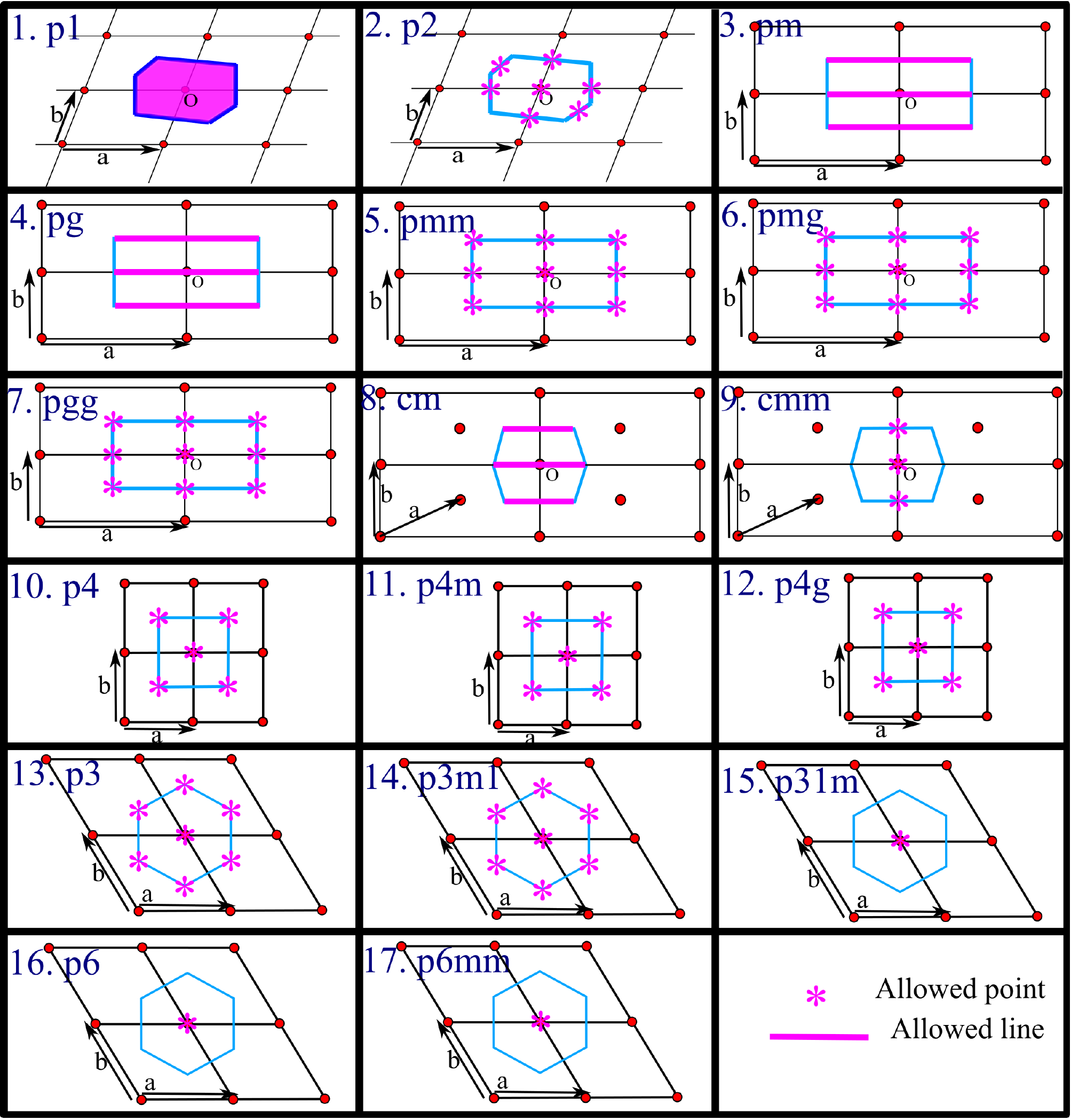}
\caption{(Color online) Allowed values of polarization for all 17 two-dimensional space groups. The corresponding Wigner Seitz cells for various lattices are displayed in blue. Allowed points and lines of points are marked in magenta. Allowed values of polarization are vectors joining the origin of the Wigner Seitz cells to the points marked.\label{fig3}}
\end{figure}

\textit{Polarization in 2D space groups.}---Similar analysis can be carried out for all 17 two-dimensional space groups.  The allowed values of polarization derived from symmetry constraints are shown in Fig.~\ref{fig3}.

Several remarks are in order.  For symmetry group $p_1$, the allowed value of $\vec P$ forms a continuum and fills up the entire Wigner Seitz cell.  This is consistent with the fact the only point group symmetry operation of $p_1$ is the identity operator and there is no topological classification protected by $p_1$.  We also note that for several symmetry groups the allowed value of $\vec P$ forms a line in one direction but is still quantized in the other direction.  In this situation, a topological classification can be obtained by taking the discrete component of $\vec P$.

\textit{Summary}---We have presented polarization as a topological index  protected by point group symmetry for crystalline insulators.  In particular, we derive 3 distinct topological classes for $C_3$ rotational symmetric systems.  Topological phase transitions between these classes are shown to be accompanied by closing of the band gap.  We complete the analysis by deriving polarization values for all 17 2D space groups which gives rich possibilities of topologically distinct classes. The topological classification scheme discussed here can be readily applied to interacting systems via the many-body formulation of polarization~\cite{martin,resta2}.

\begin{acknowledgments}
D.X. acknowledge useful discussion with Valentino R. Cooper on $h$-BN.  The authors acknowledge the allocation of computing time on NSF Teragrid machine Ranger at the Texas Advanced Computing Center. This work was supported by NRI SWAN, DOE-DMSE (DE-FG03-02ER45958), NBRPC (2012CB-921300), NSFC (91121004), and the Welch Foundation (F-1255).  D.X. was supported by the U.S. Department of Energy, Office of Basic Energy Sciences, Materials Sciences and Engineering Division.
\end{acknowledgments}

\textit{Note added.}---Upon the completion of our work, we noticed the preprint~\cite{what} on similar topics.

\end{document}